    \documentclass{aastex}
    \usepackage{graphicx}

   \shorttitle{Formation of the Solar System}
   \shortauthors{Griv \& Gedalin}

\begin{document}

   \title{The Formation of the Solar System by Gravitational Instability:
          Prediction of a new Planet or Another Kuiper-type Belt}
   \author{Evgeny Griv and Michael Gedalin}
   \affil{Department of Physics, Ben-Gurion University of the Negev,
          P.O. Box 653, Beer-Sheva 84105, Israel}

   \begin{abstract}
The early gas--dust solar nebula is considered: the gasdynamic theory is 
used to study the gravitational Jeans-type instability in its protoplanetary 
disk. The implications for the origin of the solar system are discussed. 
It is shown that a collective process, forming the basis of the gravitational
instability hypothesis, solves with surprising simplicity the two main problems
of the dynamical characteristics of the system, which are associated with
its observed spacing and orbital momentum distribution.
   \end{abstract}


   \section{Introduction}

Many young stars are surrounded by gas--dust disks (Bodenheimer \& Lin 2002).
Planetary formation is thought to start with inelastically colliding gaseous and dust 
particles settling to the central plane of a disk to form a thin and relatively dense
layer around the plane. During the early evolution of this disk it is believed that 
the dust particles coagulate into kilometer-sized rocky asteroids--``planetesimals" 
($\sim 10^{10}$ such bodies) owing to the gravitational instability (Goldreich
\& Ward 1973) and/or to the collisional sticking (Beckwith et al. 1990).
Of these processes, dust particle settling can now be observable.

We suggest that {\it all} planets 
of the solar system were created by disk instability.
That is, as a result of local gravitational instability, on attaining a certain
critical thickness (and density, respectively), small in comparison with the outer
radius of the system $R$, the circumsolar gas--dust disk disintegrated into a large 
number of separate protoplanets. 
Following Boss (2003), this hypothesis envisions coagulation and settling of dust 
grains within the protoplanets to form rock and ice cores. A protoplanet accreted
a gas subsequently from the solar nebula after accumulating a solid core of $\sim 1
\,M_\oplus$, followed by the loss of the light elements of the terrestrial planets
through the thermal emission of the sun. 
The advantages of the disk instability model are that (1) the instability process
itself is quite fast, and could form planets in $10^3-10^4$ yr (Boss 2002)
and (2) in unstable, nonaxisymmetric disks differential rotation can simultaneously
transfer angular momentum outward and mass inward through gravitational torques.
The work described here has precedents in earlier studies of gravity disturbances
in galactic disks and the Saturnian ring disk (e.g., Shu 1970; Lynden-Bell \& 
Kalnajs 1972; Griv, Yuan \& Gedalin  1999; Griv, Gedalin \& Yuan 2003).

\section{Dispersion relation}

Let us consider the dynamics of the gaseous component in the presence of the
collective self-gravitational field. A Langrangian description of the motion 
of a fluid element under the influence of a spiral field is used, looking for
time-dependent waves which propagate in a differentially rotating, 
two-dimensional disk. The approximation of an infinitesimally thin disk is a 
valid approximation if one considers perturbations with a radial wavelength that 
is greater $h$, the typical disk thickness (Toomre 1964; Shu 1970; Genkin \&
Safronov 1975; Safronov 1980).

The time dependent surface density $\Sigma (\vec{r},t)$ is splited up into a
basic and a developing (perturbation) part, $\Sigma =\Sigma_0 (r) + \Sigma_1
(\vec{r},t)$ and $|\Sigma_1/\Sigma_0|\ll 1$, where $r$, $\varphi$, $z$ are the
cylindrical coordinates and the axis of the disk rotation is taken oriented
along the $z$-axis. The gravitational potential of the disk $\aleph (\vec{r},t)$
is also of this form. These quantities $\Sigma$ and $\aleph$ are then substituted 
into the equations of motion of a fluid element, the continuity equation, the
Poisson equation, and the second order terms of the order of $\Sigma_1^2$,
$\aleph_1^2$ may be neglected with respect to the first order terms. The resultant
equations of motion are cyclic in the variables $t$ and $\varphi$, and hence by 
applying the local WKB method one may seek solutions in the form of normal modes 
by expanding any perturbation 
\begin{equation}
\Sigma_1 (\vec{r},t), \aleph_1 (\vec{r},t)= \delta \Sigma, \delta 
\aleph \, e^{i k_r r + i m \varphi - i \omega t} + \mathrm{c.\,c.} \,,
\label{eq:wkb}
\end{equation}
where $\delta \Sigma$ and $\delta \aleph$ are the real amplitudes, which are 
constant in space and time, $k_r (r)$ is the real radial wavenumber, $m$ is the
nonnegative (integer) azimuthal mode number, $\omega=\Re \omega +i\Im \omega$ is 
the complex frequency of excited waves, and $\mathrm{c.\,c.}$ means the complex
conjugate. The solution in such a form represents a spiral plane wave with $m$
arms. The imaginary part of $\omega$ corresponds to a growth ($\Im \omega >0$) or
decay ($\Im \omega <0$) of the components in time, $\Sigma_1$ and $\aleph_1 \propto 
\exp (\Im \omega t)$, and the real part to a rotation with constant angular
velocity $\Omega_{\mathrm{p}}=\Re \omega/m$. Thus, when $\Im \omega >0$, the
medium transfers its energy to the growing wave and oscillation buildup occurs.

It is important to note that in the WKB method,
the radial wavenumber is presumed to be of the form
\begin{equation}
k_r (r)={\cal{A}} \Psi (r) \,,
\label{eq:number}
\end{equation}
where ${\cal{A}}$ is a large parameter and $\Psi (r)$
is a smooth, slowly varying function of the radial distance $r$,
i.e., $\mathrm{d} \ln k_r / \mathrm{d}\ln r =O(1)$, and $|k_r| r \gg 1$.

Paralleling the analysis leading to equation (34) in Griv et al. (1999),
it is straightforward to show that 
\begin{equation}
\Sigma_1 = \frac{\beth \Sigma_0}{\omega_*^2 - \kappa^2} \left(
k_r^2 + \frac{3\Omega^2 + \omega_*^2}{\omega_*^2} \frac{m^2}{r^2} +
\frac{2\Omega}{\omega_*} \frac{m}{r} \frac{\partial \ln \Sigma_0}
{\partial r} \right) + \mathrm{c. \, c.} \,,
\label{eq:sur}
\end{equation}
where $\Sigma_1 (t \rightarrow -\infty) = 0$, so by considering only growing
perturbations we neglected the effects of the initial conditions, $\omega_*=
\omega-m\Omega$ is the Doppler-shifted (in a rotating reference frame)
wavefrequency, $\Omega (r)$ is the angular velocity of differential rotation
at the distance $r$ from the center, and $\kappa \approx \Omega$ is the epicyclic
frequency. In equation above, $\beth = \aleph_1 + P_1/\Sigma_0$,
$P_1$ is the perturbed gaseous pressure, and $c=(\partial P/\partial \Sigma)^{1/2}$ 
is the sound velocity.
In equation~(\ref{eq:sur})
only the most important low-frequency ($|\omega_*^2|\lesssim \kappa^2$) perturbations
developing in the plane $z=0$ between the inner and outer Lindblad resonances are
considered (Griv et al. 1999, 2003).
Equating the perturbed density $\Sigma_1$ [eq.~(\ref{eq:sur})] to the perturbed
density given by the asymptotic ($k_r^2 \gg m^2/r^2$) solution of the Poisson
equation (Griv et al. 1999), the Lin--Shu-type dispersion relation is obtained
\begin{equation}
\omega_{*1,2}\approx \pm p|\omega_{\mathrm{J}}| -2\pi G \Sigma_0 \frac{\Omega}
{\omega_{\mathrm{J}}^2} \frac{m}{r |k| L} \,,
\label{eq:jean}
\end{equation}
where $p=1$ for gravity-stable perturbations with $\omega_*^2 \approx
\omega_{\mathrm{J}}^2 >0$, $p=i$ for gravity-unstable perturbations
with $\omega_{\mathrm{J}}^2 < 0$, $L=(\partial \ln \Sigma_0/\partial r)^{-1}$
is the radial scale of spatial inhomogeneity, $|kL| \gg 1$, and the term
involving $L^{-1}$ is the small correction. Also,
\begin{equation}
\omega_{\mathrm{J}}^2=\kappa^2 -2\pi G\Sigma_0 (k_*^2/|k|) + k_*^2 c^2 
\label{eq:squar}
\end{equation}
is the squared Jeans frequency, $k=\sqrt{k_r^2 + m^2/r^2}$ is the
total wavenumber, $k_*^2 =k^2 \left\{ 1+[(2\Omega/\kappa)^2 -1] \sin^2 \psi
\right\}$ is the squared effective wavenumber, and $\psi = \arctan (m/rk_r)$
is the perturbation pitch angle.

Equation~(\ref{eq:jean}) determines the spectrum of oscillations. In the
gravity-unstable case, the equilibrium parameters of the disk and the 
azimuthal modeazimuthal mode number $m$ (= number of spiral arms) determine 
the spiral pattern speed of Jeans-unstable perturbations (in a rotating frame):
\begin{equation}
\Omega_{\mathrm{p}} \equiv \Re \omega_*/m \approx 2\pi G \Sigma_0 
\frac{\Omega}{|\omega_{\mathrm{J}}^2|} \frac{1}{r |k| L} \,,
\label{eq:speed}
\end{equation}
where $2\pi G \Sigma_0 |k| \sim \Omega^2$, $|\omega_{\mathrm{J}}^2| \sim
\Omega^2$, $rk^2|L| \gg 1$, and, therefore, $\Omega_{\mathrm{p}} \sim
\Omega/r k^2 L \ll \Omega$.
Thus, the typical pattern speeds of spiral structures in 
Jeans-unstable, $\omega_{\mathrm{J}}^2 < 0$, disks are only a small
fraction of some average angular velocity $\Omega_{\mathrm{av}}$.
Because $\Omega_{\mathrm{p}}$ does not depend on $m$, each Fourier
component of a perturbation in an inhomogeneous system will
rotate with the same constant angular velocity. 
The theory states that in homogeneous ($|L| \to \infty$) disks
$\Omega_{\mathrm{p}}=0$. 

The disk is Jeans-unstable to both axisymmetric (radial) and nonaxisymmetric 
(spiral) perturbations if $c < c_{\mathrm{T}}$, where $c_{\mathrm{T}} =
\pi G\Sigma_0/\kappa$ is the Safronov--Toomre (Safronov 1960, 1980; Toomre 1964) 
critical velocity dispersion to suppress the instability of 
axisymmetric ($\psi = 0$) perturbations. Thus, if the disk is 
thin, $c \ll r\Omega$, and dynamically cold, $c < c_{\mathrm{T}}$, then such a
model will be gravitationally unstable, and it should almost instanteneously (see
below for a time estimate) taken on the form of a cartwheel. The instability,
which is algebraic in nature, is driven by a strong nonresonant interaction of 
the gravity fluctuations (e.g., those produced by a spontaneous perturbation
and/or a satellite system) with the bulk of the particle population, and
the dynamics of Jeans perturbations can be characterized as a 
nonresonant interaction, that is, in equation~(\ref{eq:sur}),
$\omega_* - l \kappa \ne 0$, where $l=0,\pm 1$.

A very important feature of the instability under consideration is the
fact that it is almost aperiodic ($|\Re \omega_*/\Im \omega_*| \ll 1$).
The growth rate of the instability is relatively high:
\begin{equation}
\Im \omega_* \approx \sqrt{2\pi G\Sigma_0 (k_*^2/|k|)}
\label{eq:growthin}
\end{equation}
and in general $\Im \omega_* \sim \Omega$, that is, the instability
develops rapidly on a dynamical time scale (on a time of $3-4$ disk
rotations, or about $10^4$ yr in the early solar nebula). From
equation~(\ref{eq:squar}), the growth rate of the instability has a 
maximum at the wavelenght $\lambda_{\mathrm{crit}}\approx 2c^2/G\Sigma_0$.
At the boundary of instability, that is, $c\approx c_{\mathrm{T}}$, $\lambda_
{\mathrm{crit}} \approx 2\pi^2 G\Sigma_0 /\kappa^2 \sim 2\pi h$.
It means that of all harmonics of initial
gravity perturbation, one perturbation with $\lambda_{\mathrm{crit}}
\approx 2\pi h$, with the associated number of spiral
arms $m$, and with the pitch angle $\psi$ will be formed asymptotically 
in time af a single rotation ($\approx 5 \times 10^9$ yr ago). 
For the parameters of the solar nebula
($R \sim 300$ AU, $\kappa =2\pi/T_{\mathrm{orb}} 
\sim 10^{-10}$ s$^{-1}$, and the total mass of the disk
$M_{\mathrm{d}} \sim 0.1 \, M_\odot$), 
one obtains the typical mass of the core of a
protoplanet $M_{\mathrm{c}} \sim 10^{-6} \, M_\odot \sim M_\oplus$. 

\section{Spacing of the planets}

\begin{figure*}[htb]
\epsscale{1.0}
\plotone{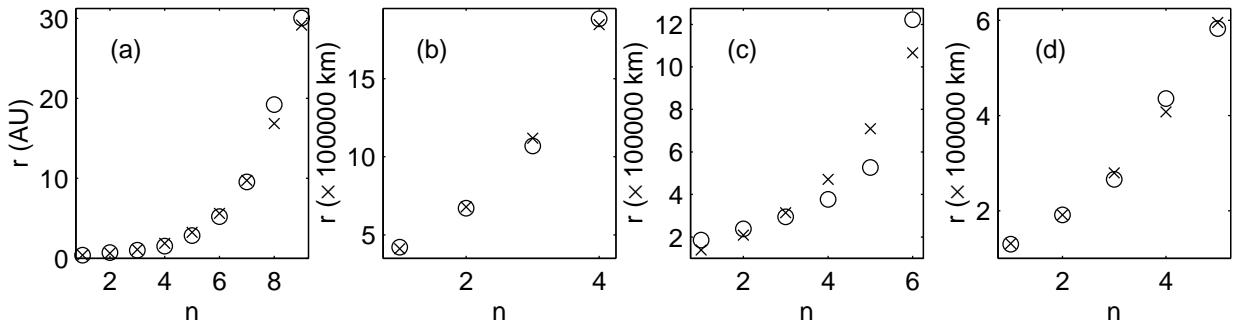}
\caption{Relation between distances of planets (satellites) from the Sun
(planets) $r$ and their numbers $n$. Data observed are represented by circles:
(a) the solar system, (b) the satellite system of Jupiter, $r_0=249.679$ and $A=1.
649$, (c) the satellite system of Saturn, $r_0=92.416$ and $A=1.503$, and (d) the
satellite system of Uranus, $r_0=89.737$ and $A=1.46$. The crosses represent the
TB rule, equation~(\ref{eq:titbod}).} 
\label{fig:titbod}
\end{figure*}

There exists the empirical Titius-Bode (TB) rule which gives the mean
orbital distances of the planets and which can be written in the
Blagg--Richardson formulation as
\begin{equation}
r_{\mathrm{n}} = r_0 A^{\mathrm{n}} \,,
\label{eq:titbod}
\end{equation}
where $r_{\mathrm{n}}$ is the distance of the nth planet from the Sun
(in AU), $\mathrm{n}=1$ for Mercury, 2 for Venus, $\dots$, and 9 for
Neptune, $A=1.73$ is the mean ratio between two consecutive planetary
distances, and $r_0 \approx 0.21$. Also, one cannot overlook the fact that
many of the regularities which are found in the planetary system are also to 
be seen in the regular satellite systems of Jupiter, Saturn, and Uranus, e.g.,
the spacing of the regular satellites is a variation of the TB rule
(Fig.~\ref{fig:titbod}). This suggests that the same cosmogonic process must
have been responsible for the origin of both types of systems. 
Lynch (2003) has already
argued that it is not possible to conclude unequivocally that laws of TB
type are, or are not, significant. Therefore, the possibility of a physical
explanation for the observed distributions remains open.\footnote{
Interestingly, the mean orbital distance to the recently discovered classical
Edgeworth--Kuiper belt objects, $r \approx 46$ AU, is in fair agreement with
that given by the TB rule for the solar system's 9th planet, 
$r_{\mathrm{10}}\approx 50$ AU.}

\begin{figure*}[htb]
\epsscale{1.0}
\plotone{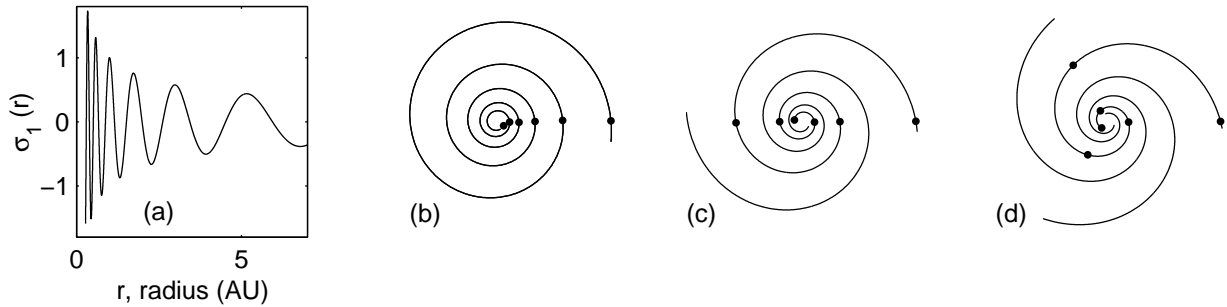}
\caption{(a) Dependence of the perturbed surface density of the protoplanetary
disk $\Sigma_1 (r)$ (arbitrary units) on the radius $r$, equation~(\ref{eq:cos}). 
The maxima of the perturbed density coincide with locations of all the planets.
(b) Spiral density waves with $m=1$ arm [eq.~(\ref{eq:cos})] in the ($r,
\varphi$)-plane, (c) density waves with $m=2$ arms, and (d) density waves with
$m=3$ arms. The filled circles represent the maxima of the perturbed density
(protoplanets) of Jeans waves,
which are unstable to both axisymmetric and nonaxisymmetric perturbations.}
\label{fig:positions}
\end{figure*}

Equation~(\ref{eq:titbod}) can be rewritten:
\begin{equation}
(2 \pi /\ln 1.73) \ln (r_{\mathrm{n}}/0.21) = 2\pi \mathrm{n} \, .
\label{eq:titbod2}
\end{equation}
Next, the surface density of the disk may be represented in the form of 
the sum of the equilibrium surface density $\Sigma_0 (r)$ and the perturbed 
surface density
\begin{equation}
\Sigma_1 (r)=\delta \Sigma (r) e^{\Im \omega t} \cos \left[ 11.5
\ln (r_{\mathrm{n}}/0.21) + m \varphi \right] \, ,
\label{eq:cos}
\end{equation}
where $\delta \Sigma (r)$ is the amplitude 
varying slowly with radius, and $\left[ 11.5 \ln (r_{\mathrm{n}}/0.21)
+ m \varphi \right]$ represents the phase varying rapidly
with radius,
\begin{displaymath}
|k_r|r \equiv 11.5 \left| (\mathrm{d}/\mathrm{d}r) \ln
(r_{\mathrm{n}}/0.21) \right| r \gg 1 \, .
\end{displaymath}
Equation~(\ref{eq:titbod2}) and the condition $\delta \Sigma (r)>0$ on
the initial phase imply that the maximum values of the perturbed density
in equation~(\ref{eq:cos}) coincide with the positions of all the planets
(Fig.~\ref{fig:positions}a).

Interestingly, and this is the central part of our theory, the TB rule
[eq.~(\ref{eq:cos})] {\it satisfies} the conditions of the WKB wave
with the effective TB radial wavenumber
\begin{equation}
k_{\mathrm{eff}} = \frac{11.5}{r_{\mathrm{n}}} \, ,
\label{eq:wkbtb}
\end{equation}
$\mathrm{d} \ln k_{\mathrm{eff}}/\mathrm{d}\ln r =O(1)$, and $k_
{\mathrm{eff}}r \gg 1$ [cf. eq.~(\ref{eq:number})].\footnote{ 
Polyachenko (Polyachenko \& Fridman 1972) has already been considered this
analogy in his investigation of the possibility of the explanation of
the law of planetary distances by the gravitational instability in
sufficiently flat systems, but evidently without success. In particular,
Polyachenko studied only axisymmetric $m=0$ perturbations, which do not
carry angular momentum (see an explanation below in $\S 4$).}

Thus, if the space dependence of the perturbed surface density of 
the protoplanetary disk in the ($r,\varphi$)-plane has the form of
equation~(\ref{eq:cos}) with $\Im \omega > 0$, the maxima of both
radially and azimuthally unstable gravity perturbations are located
in places of the solar system's planets (Figs~\ref{fig:positions}b, c, d).
Let us define conditions under which the density maxima are localized on 
planetary orbits. If the disk is inhomogeneous with respect to equilibrium
parameters, the wavelength of a perturbation with a maximum growth rate
$\lambda_{\mathrm{crit}}$ will be a function of the radius $r$. From the 
above, the wavelength $\lambda_{\mathrm{crit}}\approx 4\pi^2 G\Sigma_0 
/\kappa^2$, corresponding to the minimum on the dispersion 
curve~(\ref{eq:jean}) (see also Griv et al. 2003, Fig. 1 therein). On the
the other hand, the wavelength is $\lambda_{\mathrm{eff}}=2\pi / k_
{\mathrm{eff}}$. Comparing $\lambda_{\mathrm{crit}}$ with $\lambda_
{\mathrm{eff}}$, we see that in the case where the disk density is
dependent on radius according to the law
\begin{equation}
\Sigma_0 (r) \approx 0.0138 G^{-1} \kappa^2 r \,,
\label{eq:law}
\end{equation}
the maxima of time-increasing, both radially and azimuthally Jeans-unstable 
density perturbations are arranged in it according the TB rule given by
equation~(\ref{eq:titbod}). 
The last condition may be fulfilled in Keplerian disks, $\kappa \propto
r^{-3/2}$, only if $\Sigma_0 \propto r^{-2}$. Interestingly, Tomley et al. (1991)
have used almost the same law $\Sigma_0 \propto r^{-7/4}$ as initial profile 
for simulation of a disk surrounding the central star. The reason for using such
a law comes from a particular model of protostellar cloud collapse Tomley et al.
used. It was obtained that this initial model did not get much subsequent
evolution in the simulations although it was nicely gravitationally unstable.
Based on hydrostatic models, the radial density distributions in circumstellar 
disks around Herbig Ae/Be and T Tauri stars have been proposed to be in the 
range $\Sigma_0 \propto r^{-(1.9-2.4)}$. Detailed modeling of the
NIR-to-millimeter appearance of several spatially resolved T Tauri disks has
confirmed these predictions. It has been stated that optically thick young
disks around those stars with spatial structures are dominated by gravitation
and gasdynamics. See, e.g., Stapelfeldt et al. (1998), Chiang \& Goldreich
(1999), and Wolf, Padgett \& Stapelfeldt (2003) for a discussion. Fits of
models to observed spectral energy distributions of protostellar disks
typically give $\Sigma_0 \propto r^{-3/2}$ (Bodenheimer \& Lin 2002).
Also, a standard reference model of a disk, known as the 
``minimum mass solar nebula," reconstructed from the distribution of mass
in the planets of the solar system and assuming solar composition and no
migration of planets, gives $\Sigma_0 \propto r^{-3/2}$. The latter is close
to the $r^{-2}$ distribution advocated above. Clearly, given the observational
and analytical uncertainties, the two distributions, $\Sigma_0
\propto r^{-3/2}$ and $\Sigma_0 \propto r^{-2}$, are not necessarily 
inconsistemt with each other. For instance, the inclusion of the disk's
self-gravity in addition to the gravitational field of the sun will reduce 
the value of the exponent $n$ in the required density--radius relation $\Sigma_0 
\propto r^{-n}$. In turn, both optical and near-infrared observations of
pre-main-sequence stars of intermediate mass have revealed the spiral
structure, and thus presumably the Jeans instability, in the circumstellar
disks (Grady et al. 1999; Clampin et al. 2003; Fukagawa et al. 2004).

One concludes, therefore, that if the surface density of a protoplanetary
disk falls according to the law given by equation~(\ref{eq:law}),
the increasing maxima of density perturbations of a Safronov--Toomre-unstable
disk ($c < c_{\mathrm{T}}$) are located between the Lindblad resonances in
places of the planets (Fig.~\ref{fig:positions}).
We believe to have obtained a
theoretical interpretation of the TB rule: the distance between planets is 
the wavelength of the most Jeans-unstable perturbations at the given 
point of the protoplanetary disk.

By using equation~(\ref{eq:law}), it is easy to find that the disk
mass between 0.3 AU and 30 AU is about $0.4 \, M_\odot$. This means
that in the present planets there is contained not more than about
$0.5\%$ of the mass of the protoplanet cloud. Almost certainly, a part 
of the initial mass of the planets was blown away due to intensive
corpuscular emission of the early sun.

\section{Orbital momentum distribution}

We next turn to the question of how to account for the concentration 
of angular momentum in the planets and of mass in the sun. The torque
exerted by the gravity perturbations on the disk is ${\cal{T}}=-\int
\int \mathrm{d}^2 r (\vec{r} \times \vec{\nabla} \aleph_1) \Sigma_1$ or
\begin{equation}
{\cal{T}} =-\int_{r_1}^{r_2} r \mathrm{d} r \int_0^{2\pi} \Sigma_1
(r,\varphi^\prime) \frac{\partial \aleph_1 (r,\varphi^\prime)}
{\partial \varphi^\prime} \mathrm{d} \varphi^\prime \,.
\label{eq:momen}
\end{equation}
The points $r_1$ and $r_2$ in which $\omega_* \pm \kappa = 0$ are
called the points of inner and outer Lindblad resonances. They play an
important role in the theory: the solution of spiral type~(\ref{eq:wkb})
rapidly oscillating in the radial direction lies between $r_1$ and $r_2$.
Outside the resonances, $r<r_1$ and $r>r_2$, the solution decreases
exponentially. A special analysis of the solution near corotation ($
\omega_* = 0$) and Lindblad resonances is required. Resonances of a higher
order, $\omega_* \pm l\kappa = 0$ and $|l| =2, 3, \cdots$, are dynamically
of less importance (Shu 1970). To emphasize it again, the 
present analysis is restricted to consideration of only the principal part
of a disk between the Lindblad resonances. 
Investigation of the wave--particle interaction at spatially limited
resonances has been done by Lynden-Bell \& Kalnajs (1972), Goldreich \&
Tremaine (1978, 1980), and Griv, Gedalin, Eichler \& Yuan (2000).

Using equation~(\ref{eq:sur}), from equation~(\ref{eq:momen}) one finds
\begin{equation}
{\cal{T}} \approx - \frac{8\pi \Sigma_0}{\Omega \Im \omega_*} 
m^2 \aleph_1 \aleph_1^* \,,
\label{eq:fin}
\end{equation}
where $\Im \omega_* > 0$, $\aleph_1^*$ is the complex conjugate potential, 
and the values of $\aleph_1$, $\aleph_1^*$, $\Sigma_0$, $\Omega$ are evaluated 
at $r=r_1$. Three physical conclusions can be deduced from 
equation~(\ref{eq:fin}). First, the distribution of the angular momentum 
of a disk will change under the action of only the nonaxisymmetric
forces $\propto m$. The latter is obvious: axially symmetrical
motions of a system, studied by Polyachenko, produce no
gravitational couples between the inner parts and the outer parts. Second,
the distribution of the angular momentum will change upon time only
under the action of growing, i.e., Jeans-unstable perturbations ($\Im 
\omega_* > 0$).\footnote{In the opposite limiting case of slow growth ($\Im
\omega_* \rightarrow 0$), absorption and emission of angular momentum are
confined only to resonate particles (e.g., Lynden-Bell \& Kalnajs 1972). 
The treatment of resonances is beyond the scope of the 
present analysis.}
Third, ${\cal{T}} < 0$: the spiral
perturbations remove angular momentum from the disk. This takes place 
in the main part of the disk between the Lindblad resonances where
spiral density waves are self-excited via a nonresonant wave--``fluid"
interaction. Further there is absorption of angular momentum by particles 
that resonate with the wave (Lynden-Bell \& Kalnajs 1972).
As a result, the bulk of angular momentum is transferred outward (and a
mass transported inward, correspondingly). In turn a small group of resonate
particles moves outward taking almost all angular momentum.\footnote{
Lynden-Bell \& Kalnajs (1972) have proved that in good conformity
with $N$-body simulations the gravitational torques can only communicate
angular momentum outward if the spirals trails.} 
These processes lead to the core-dominated mass density profile in the 
protoplanetary disk, together with the buildup of an extended, rapidly
rotating outer envelope. We speculate that a large portion of the initial 
mass of the nebula was transported toward the sun. 

Let us evaluate the gravitational torque for a realistic model of the
protoplanetary disk. In accordance with the theory developed above, the 
fastest growing spiral mode with $m\gtrsim 1$, $k_*=k_{\mathrm{crit}}$, and 
$\Im \omega_* \sim \Omega$ is considered. Taking into account that $8\pi m^2 
\aleph_1 \aleph_1^* \sim \aleph_0^2$ (an astrophysicist might well consider a
perturbation with $\aleph_1/\aleph_0$ of 1/10 or even 1/3 to be quite small)
and $\aleph_0 \sim r^2 \Omega^2$, where $\aleph_0$ is
the basic potential, from equation~(\ref{eq:fin}) one obtains
$|{\cal{T}}| \sim \Sigma_0 r^4 \Omega^2$. The angular momentum of the 
disk ${\cal{L}} \sim \Sigma_0 r^4 \Omega$. Then the characteristic time of
the angular momentum redistribution is $t \sim {\cal{L}}/{\cal{T}} \sim
\Omega^{-1}$.
Thus, already in the first $3-4$ disk revolutions, in say about $10^4$ yr,
the gas--dust protoplanetary disk sees its almost all angular momentum
transferred outward and mass inward. We conclude that the  
Jeans instability studied here can give rise to
torques that can help to clear the nebula on a time scale of
$\gtrsim 1$ Myr, in accord with astronomical requirements. In addition,
the analysis is found to imply the existence of a {\it new} planet (or 
another Kuiper-type belt) at a mean distance from the sun of $r_{\mathrm{11}} 
= 0.21 \times 1.73^{11} \approx 87$ AU.

\end{document}